\documentclass[fleqn,usenatbib]{mnras}
\usepackage{newtxtext,newtxmath}
\usepackage[T1]{fontenc}
\DeclareRobustCommand{\VAN}[3]{#2}
\let\VANthebibliography\thebibliography
\def\thebibliography{\DeclareRobustCommand{\VAN}[3]{##3}\VANthebibliography}
\usepackage{graphicx} 
\usepackage{amsmath}	
\usepackage{subcaption}

\def\flx{erg~cm$^{-2}$~s$^{-1}$}
\def\lum{erg~s$^{-1}$}

\def\arcsec{\hbox{$^{\prime\prime}$}}

\def\approxlt{\ifmmode \rlap{$<$}{}_{{}_{{}_{\textstyle\sim}}} \else%
$\rlap{$<$}{}_{{}_{{}_{\textstyle\sim}}}$\fi}

\title[EP250207b host galaxy at z=3.2]{A JWST redshift for the host galaxy of EP250207b of $z=3.2$: a collapsar origin is viable
}

\author[van Hoof et al.]{Agnes P. C. van Hoof,$^{1}$
          Peter G. Jonker,$^{1}$
          Andrew J. Levan,$^{1,2}$
          Nial R. Tanvir,$^{3}$
          Franz E. Bauer,$^{4}$ 
          Joe Bright,$^{5,6}$          
          \newauthor
          Francesco Carotenuto,$^{7}$
          Ting-Wan Chen,$^{8}$
          Ashley Chrimes,$^{9,1}$
          Gregory Corcoran,$^{10}$
          Laura Cotter,$^{10}$
          \newauthor
          Joyce N. D. van Dalen,$^{1}$
          Rob A.J. Eyles-Ferris,$^{3}$
          Morgan Fraser,$^{10}$
          Daniele B. Malesani,$^{11,12,1}$
          \newauthor
          Daniel Mata S\'anchez,$^{13,14}$
          Antonio Martin-Carrillo,$^{10}$
          Paul O’Brien,$^{3}$
          Francesca Onori,$^{7}$
          \newauthor
          Jonathan Quirola-V\'asquez,$^{1}$
          Maria E. Ravasio,$^{15,16,1,17}$
          Andrea Rossi,$^{18}$
          Javi S\'anchez-Sierras,$^{1}$
          \newauthor
          Nikhil Sarin,$^{19,20}$
          Steve Schulze,$^{21}$
          Hui Sun,$^{22}$
          Manuel A.P. Torres$^{13,14}$
\\
   $^{1}$Department of Astrophysics/IMAPP, Radboud University, 6525 AJ Nijmegen, The Netherlands \\
   $^{2}$Department of Physics, University of Warwick, Coventry, CV4 7AL, UK \\
   $^{3}$School of Physics and Astronomy, University of Leicester, University Road, LE1 7RH, UK \\
   $^{4}$Instituto de Alta Investigaci\'on, Universidad de Tarapac\'{a}, Casilla 7D, Arica, Chile \\
   $^{5}$Department of Physics, University of Oxford, Keble Road, Oxford, OX1 3RH, UK \\ 
   $^{6}$Breakthrough Listen, Astrophysics, Department of Physics, The University of Oxford, Keble Road, Oxford OX1 3RH, UK \\
   $^{7}$INAF-Osservatorio Astronomico di Roma, Via Frascati 33, I-00078, Monte Porzio Catone, Italy\\
   $^{8}$Graduate Institute of Astronomy, National Central University, 300 Jhongda Road, 32001 Jhongli, Taiwan \\
   $^{9}$European Space Agency (ESA), European Space Research and Technology Centre (ESTEC), Keplerlaan 1, 2201 AZ Noordwijk, The Netherlands \\
   $^{10}$School of Physics and Centre for Space Research, University College Dublin, Belfield, Dublin 4, Ireland \\
   $^{11}$Cosmic Dawn Center (DAWN), Denmark\\
   $^{12}$Niels Bohr Institute, University of Copenhagen, Jagtvej 128, Copenhagen, 2200, Denmark\\
   $^{13}$Instituto de Astrof\'isica de Canarias, E-38205 La Laguna, Tenerife, Spain \\
   $^{14}$Departamento de Astrof\'isica, Univ. de La Laguna, E-38206 La Laguna, Tenerife, Spain \\
   $^{15}$Institute of Space Sciences (ICE, CSIC), Campus UAB, Carrer de Can Magrans s/n, Barcelona, E-08193, Spain \\
    $^{16}$Institut d’Estudis Espacials de Catalunya (IEEC), Edifici RDIT, Campus UPC, Castelldefels (Barcelona), E-08860, Spain \\
    $^{17}$INAF – Osservatorio Astronomico di Brera, via Emilio Bianchi 46, I-23807 Merate (LC), Italy \\
   $^{18}$INAF–Osservatorio di Astroﬁsica e Scienza dello Spazio, via Piero Gobetti 93/3, I-40129 Bologna, Italy \\
   $^{19}$Kavli Institute for Cosmology Cambridge, Madingley Road, Cambridge CB3 0HA, United Kingdom, \\
   $^{20}$Institute of Astronomy, University of Cambridge, Madingley Road, Cambridge CB3 0HA, United Kingdom \\
   $^{21}$Department of Particle Physics and Astrophysics, Weizmann Institute of Science, 234 Herzl St, 76100 Rehovot, Israel \\
   $^{22}$National Astronomical Observatories, Chinese Academy of Sciences, Beijing 100101, China
}

\date{Accepted XXX. Received YYY; in original form ZZZ}

\pubyear{\the\year{}}

\begin{document}
\label{firstpage}
\pagerange{\pageref{firstpage}--\pageref{lastpage}}
\maketitle

\begin{abstract}
We present James Webb Space Telescope (JWST) and Hubble Space Telescope (HST) observations of the field of the fast X-ray transient (FXT) detected by Einstein Probe, EP250207b, to resolve any ambiguity about the host galaxy and redshift of the FXT. EP250207b was originally associated with a nearby galaxy at $z=0.082$, based on its low chance alignment probability, and a binary neutron star merger origin was proposed. However, we report the detection of a background galaxy at $z=3.2$ at the location of EP250207b. Assuming this galaxy is the actual host galaxy, the rest-frame energetics and timescales of the event change. Furthermore, the available data are not able to rule out the presence of a supernova associated with EP250207b if at this redshift. We model the X-ray, optical, near-infrared and radio light curves using a tophat jet model implemented in \textsc{Redback} and find that they are consistent with an on-axis gamma ray burst afterglow. The energetics and host galaxy properties do not allow us to distinguish between a collapsar and a merger driven event. 
\end{abstract}


\section{Introduction}
Several of the extra-galactic fast X-ray transients (FXTs) discovered by the Einstein Probe (EP) satellite \citep{Yuan2022} have been associated with gamma-ray bursts (GRBs). These associations are established by the contemporaneous detections by EP and gamma-ray satellites \citep[e.g.~][]{Yin_2024, jiang2025} or via the emergence of a broad-lined type Ic supernova \citep[Ic-BL SN; e.g.,][]{vandalen2025, rastinejad2025, quirolavasquez2026, vanhoof2026} similar to those that have been found to accompany low redshift collapsar GRBs \citep[initially referred to as long GRBs; e.g.~][and used interchangeably here]{Rastinejad2022}. \cite{OConnor2025} additionally show that EP discovered FXTs seem to have a similar redshift distribution as collapsar GRBs.
However, to date there has not been much incontrovertible evidence for associations with compact object mergers, such as binary neutron star (BNS) mergers that power short GRBs, except for the case of EP250704a which had a coinciding short GRB \citep[][]{fraija2026, Li2026}. 
Nonetheless, for many FXTs discovered by EP, no multi-wavelength counterpart was found, making it much more challenging to establish or rule out an association with specific progenitor models. In these cases, or in cases where no spectroscopic redshift could be obtained for the counterpart, the association to a candidate host galaxy using mainly the statistics of the sky distribution of galaxies and their magnitudes (e.g.~\citealt{Bloom2002}) can provide tentative constraints to the nature of the FXT \citep[e.g.,][]{Lin2022, Eappachen_2022, Quirola_Vasquez_2022,Quirola_Vasquez_2023, Brightman2026, vanhoof2026b}. However, these associations are less certain than those where a spectroscopic redshift measurement is available for both the host galaxy and the counterpart of the event \citep[e.g.,][]{quirolavasquez2025}, particularly when the underlying host distribution is unknown.

Recently, \cite{Becerra2026} and \cite{Jonker2026} reported the discovery and multi-wavelength analysis of the FXT EP250207b. Based on the projected offset to and the bright apparent magnitude of the candidate host, there is a low chance alignment probability \citep{Bloom2002} of P$_{\rm chance} \lesssim 0.5$~\%. Hence, \cite{Becerra2026} and \cite{Jonker2026} proposed its association with a galaxy at redshift $z=0.082$, while noting that the lack of spectroscopic redshift of the counterpart does not rule out other scenarios. If at $z=0.082$ the event is most consistent with a compact object merger. If confirmed this would have been the second potential link between FXTs and compact object mergers, following the case of GRB~250704B/EP250704a \citep{fraija2026,Li2026}. These authors discussed the possibility that there is an unresolved background galaxy at higher redshift with P$_{\rm chance} \approx 10$~\% as calculated by \cite{Jonker2026} using the \cite{Bloom2002} formula and P$_{\rm chance} \approx 0.6$~\% as calculated by \cite{Becerra2026} using the half-light radius and the surface density of the galaxy. 

In this Letter, we present \emph{James Webb Space Telescope} (JWST) and \emph{Hubble Space Telescope} (HST) observations of the field of EP250207b in order to help resolve the uncertainty about the host galaxy and thus the redshift, energetics and physical origin of this FXT. We present these observations and our results in Section~\ref{obs_res}, discuss the results in Section~\ref{discussion} and conclude in Section~\ref{conclusion}.
All magnitudes throughout this work are reported in the AB magnitude system and we assume the $\Lambda$CDM Planck cosmology \citep{2020A&A...641A...6P} with $H_0$~=~67.7~km~s$^{-1}$~Mpc$^{-1}$ and $\Omega_m$~=~0.31.

\section{Observations \& Results}
\label{obs_res}
\begin{figure*}
    \centering
    \begin{subfigure}{0.49\textwidth}
    \centering
    \includegraphics[width=0.95\linewidth]{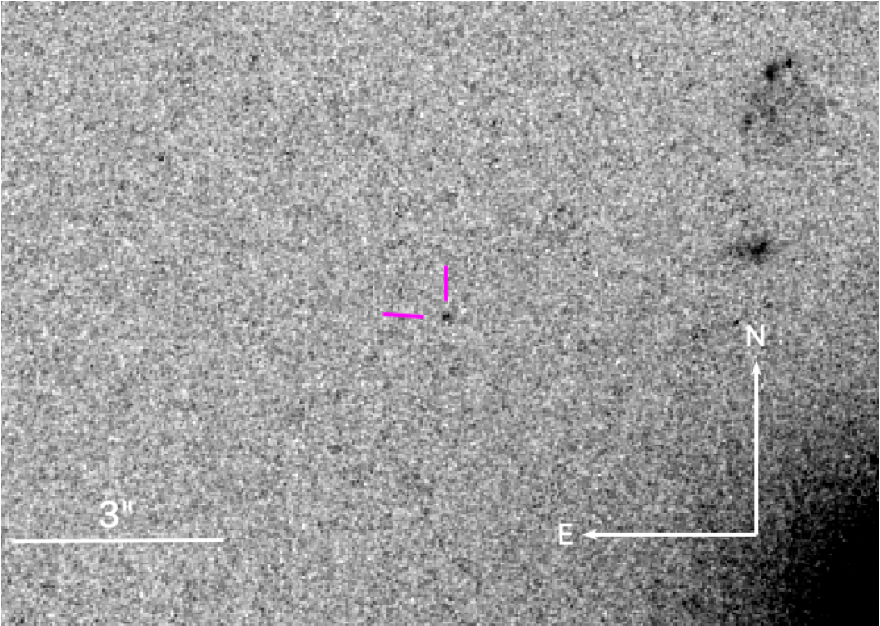}
    \label{Fig:Finderf606w}
    \end{subfigure}
    \centering
    \begin{subfigure}{0.49\textwidth}
    \centering
    \includegraphics[width=0.95\linewidth]{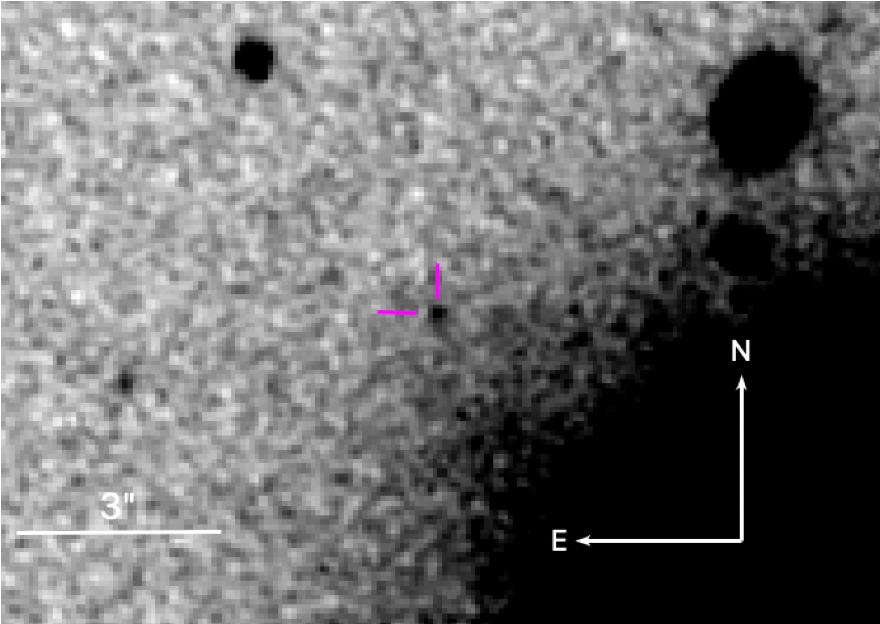}
    \label{Fig:Finderf160w}
    \end{subfigure}
    \centering
    \begin{subfigure}{\textwidth}
    \centering
	\includegraphics[width=\textwidth]{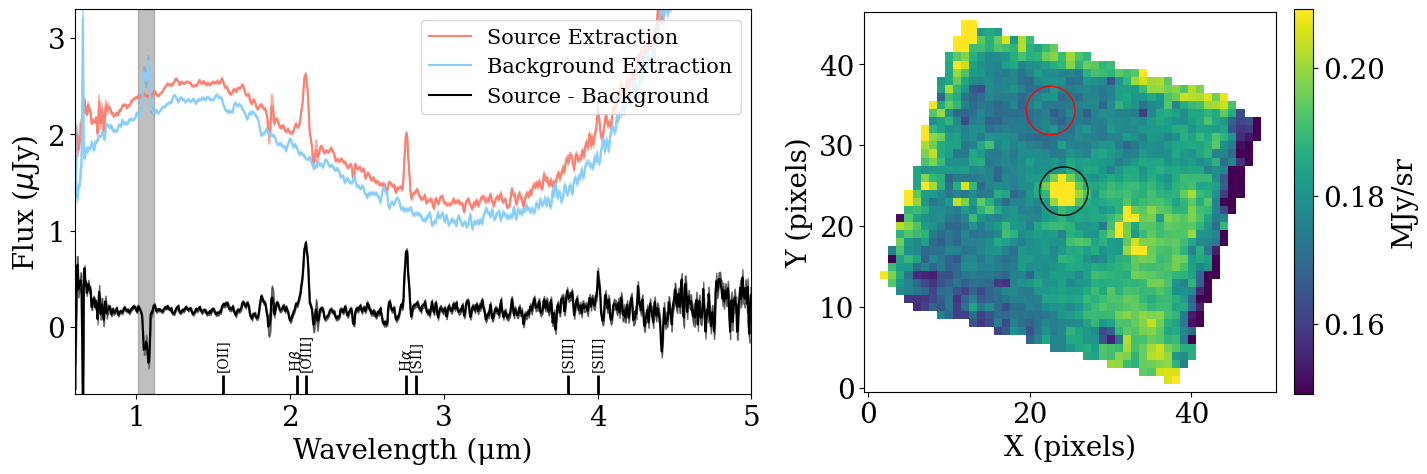}
    \end{subfigure}
        \caption{ \textit{Top:} HST images of the field of EP250207b, $\sim$289 days after the EP-WXT trigger. A source is clearly visible in both the F606W (left) and the F160W (right) image, as indicated by the magenta markers.
        \textit{Bottom left:} Extracted 1D spectrum of the JWST/NIRSpec IFU data in observed wavelength ($\mu$m) and flux density in $\mu$Jy. Extraction at the source position is shown in light red, the background region in blue and the background subtracted source in black. The locations of several emission lines at $z=3.2$ are over shown with vertical lines. The grey shaded region denotes a region of high noise in the background, which is masked. \textit{Bottom right:} Collapsed IFU Cube with the extraction aperture shown in black and the background region with in red. The axes are pixels and the colour indicates the flux density in MJy/sr.}
        \label{fig:IFU}
\end{figure*}

We observed the field of the optical counterpart of EP250207b \citep{Jonker2026,Becerra2026} with HST on 2025 Nov 23 ($\sim$289 days after the EP Wide-field X-ray Telescope (EP-WXT) trigger) using the Wide Field Camera 3 (WFC3) in the filters F160W and F606W. These observations are part of program \#17806 with PI Tanvir. We obtained $4\times505$s of observations in F606W and $4\times552.94$s in F160W. The images are drizzled to pixel scales of 0.05\arcsec~and 0.07\arcsec~in F606W and F160W, respectively. The images are shown in the top panels of Figure~\ref{fig:IFU}.
We detect a source at the transient position at magnitudes $m_{\rm F606W}~=~26.74~\pm~0.16$ and $m_{\rm F160W}$~=~27.11~$\pm$~0.10. For the F606W filter, this is consistent with the last epoch (at $t=29$~d in the observer frame) presented in \cite{Jonker2026} when a magnitude of $m_{\rm F606W}$~=~26.86~$\pm$~0.15 was measured. The magnitude in the F160W NIR filter at that epoch was $m_{\rm F160W}$~=~26.32~$\pm$~0.18, indicating that the source faded significantly between that epoch and the observations we present here.
Using Source Extractor \citep[SE;][]{sextractor} we find an ellipticity of $0.22~\pm~0.13$, as measured on the F606W image and $0.4\pm0.1$ from the F160W image, showing that the source is marginally elongated at least in the F160W band. The full width at half maximum (FWHM) of the source, as determined using the \texttt{FWHM\_IMAGE} argument of SE on the F606W image, is 4.4~pixels or 0.15\arcsec, on the F160W image, we measure \texttt{FWHM\_IMAGE}$=5.7$~pixels or 0.4\arcsec. They are both larger than the diffraction limit of HST in the corresponding filters, implying that the source is resolved in both images, as could be the case for a (host) galaxy.

We observed this galaxy with the JWST NIRSpec integral field unit (IFU) using the prism and an exposure time of 3151~s on 2025 Dec 22 with program \#06970, PI Jonker. We extract the spectrum using the \texttt{Extract1DStep} of the NIRSpec-IFU JWST pipeline, centred on the galaxy with an 0.3\arcsec~aperture, as indicated with the black circle in the bottom-right panel of Fig.~\ref{fig:IFU}. We subtract the background from an aperture of the same size to the North-East of the target, rather than from an annulus around the source to have as little interference of light and potential spectral features from the large foreground galaxy as possible, as indicated with the red circle in the bottom-right panel of Fig.~\ref{fig:IFU}. As this background region still shows a feature between 1.01 and 1.12 $\mu$m, some features appear in the background-subtracted source spectrum and we mask this region for the rest of our analysis. The resulting 1D spectra and the masked region are shown in the bottom-left panel of Fig.~\ref{fig:IFU}.

In the spectrum, we detect multiple clear emission lines, consistent with H$\alpha$, H$\beta$, [OIII], [SII] and [SIII] at $z=3.1999~\pm~0.002$ ($z=3.2$, hereafter). We note that the H$\beta$ and [OIII] lines are blended, so we fit two Gaussians to the blended line. We first normalise the continuum and then measure the equivalent width (EW) of the lines. We find that the EW of H$\alpha$ is $\sim$3 times that of H$\beta$, as expected. Hence, the features are indeed blended.

\section{Discussion}
\label{discussion}
We found a galaxy at redshift $z=3.2$ at the position of the counterpart of the FXT EP250207b. Based on a chance alignment probability of $P_{chance}\approx0.06$\% \citep[calculated from the F606W magnitude and formula in][we note that this formula is based on a galaxy distribution for lower redshift galaxies]{Bloom2002}, we assume that this is the host galaxy of EP250207b.

\subsection{Energetics}
The average unabsorbed EP-WXT 0.5--4~keV X-ray flux of EP250207b of $(6.5~\pm~3.6)~\times~10^{-10}$~\flx~\citep[averaged over 120~s;][]{Jonker2026} 
corresponds to a luminosity of L$_{X}\approx3\times10^{49}$~\lum~at $z=3.2$. This X-ray luminosity is in line with that of collapsar GRBs \citep[e.g.~][]{margutti2011}, but it is too high to be explained by the spin-down energy of a millisecond magnetar formed after a BNS merger \citep[e.g.,][]{Metzger2008, Siegel2016}. It can, however, also be powered by the X-ray emission of the jet during the prompt emission in merger-driven short GRBs \citep[e.g.~][]{Ierardi2026}. We also re-calculate the upper limits on the radio luminosity from the MeerKAT non-detections at $t$~=~5.6, 23, and 43 days from the flux levels of $\sim$25$\mu$Jy/beam \citep{Jonker2026} and an additional epoch at 85~days, to be L$_{3.06 GHz} \lesssim 7.2 \times 10^{40}$~\lum, consistent with the radio luminosity of short GRB afterglows as shown in Fig.~\ref{fig:radio}. It is also in line with that of a large fraction of the long GRB population \citep[][and references therein]{Fong_2021,Margutti2014}.

\subsection{Light curve \& modelling}
At the revised redshift of $z=3.2$, the peak $r^{\prime}$-band magnitude reported by \cite{Jonker2026} corresponds to a peak absolute magnitude of $M_{r^{\prime}}$~=~$-22.3$. This is consistent with the peak absolute magnitude of several other EP-discovered FXTs and the population of long GRBs \citep[e.g.~][]{Kann2006}, while it is on the bright end of the observed distribution for short GRBs \citep[e.g.~][]{NicuesaGuelbenzu2012}. 
Only one epoch of the reported HST observations of the counterpart of EP250207b probes the phase at which a type Ic-BL SN is expected to contribute significantly to the total emission in addition to a contribution of the afterglow, i.e.~the epoch at $\sim$~29 days observer frame ($\sim$~7 days rest frame). We measure the magnitude of the host galaxy in the latest HST epochs in order to assess the host galaxy contamination to the photometry reported in \cite{Jonker2026}. For F606W, the measured magnitude is consistent with the second epoch while the magnitude in the last epoch in the F160W band we report here is fainter than that reported in \cite{Jonker2026}, indicating that there was still some contribution to the F160W emission by the transient in the second epoch observation. Strong line blanketing in Ic-BL SNe suppresses emission blueward of $3000$~\AA. This implies that any SN signature at $z=3.2$ would only be detectable in filters redward of $\sim$13000~\AA. From the observations reported here and in \cite{Jonker2026}, this includes only F160W. From our photometry, within the 1$\sigma$ uncertainty, there is the possibility that the fading between the two epochs in F160W is less than in the bluer filters, which could potentially be explained by an emerging SN component. However, the limited temporal coverage makes that we are unable to confirm or rule-out the emergence of a type Ic-BL SN in the light curve of EP250207b. 

\begin{figure}
    \includegraphics[width=\linewidth]{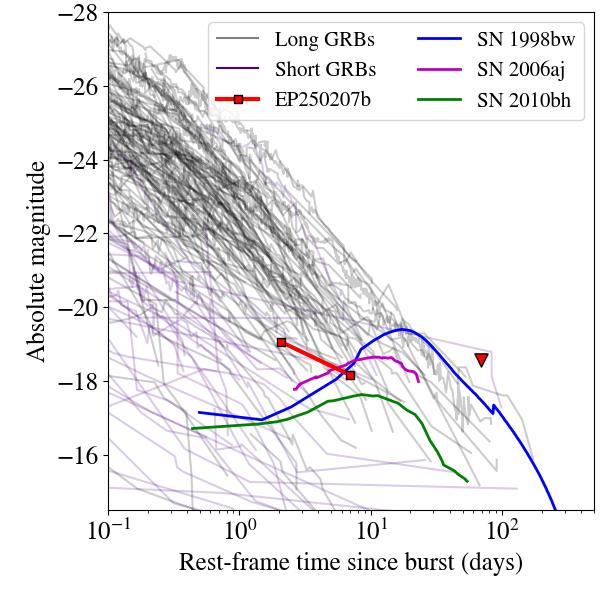}
    \caption{Kann plot showing the absolute $r^{\prime}$-band light curves of a sample of long GRBs (black), a sample of short GRBs (purple) and three well-studied GRB-SNe: SN~1998bw (blue), SN~2006aj (magenta) and SN~2010bh (green). We overplot the host-subtracted F160W observations of EP250207b in red and the host magnitude as upper limit for epoch 3. The light curve of EP250207b is consistent with that of faint long GRBs but also with that of bright short GRBs. The peaks of the SNe are just after the second epoch and fainter than the third epoch of EP250207b, leaving us unable to rule out or confirm an association with a similar SN.}
    \label{fig:Kann}
\end{figure}

We compare the afterglow light curve in absolute magnitude of the F160W band, which falls close to rest-frame $r^\prime$-band, and rest-frame time to that of the population of long and short GRBs in a Kann plot shown in Fig.~\ref{fig:Kann}. We show 135 light curves of long GRBs, obtained from \cite{Kann2006} and 40 light curves of short GRBs, obtained from \cite{NicuesaGuelbenzu2012}.
The magnitude of the host galaxy is subtracted from the first two epochs and used as an upper limit for the transient in the third epoch. We can see that the absolute magnitudes of the light curve are consistent with that of long GRBs and with the bright end of short GRBs. We also plotted the light curves of three type Ic-BL SNe associated with long GRBs/X-ray flashes, SN~1998bw, SN~2006aj and SN~2010bh \citep[][respectively]{Galama1998, Mirabal2006, Cano2011}. The SNe light curves indeed reach their peak after the second epoch of observations of EP250207b and the upper limit of epoch three is not deep enough to rule out an SN association.  

\begin{figure}
    \centering
    \includegraphics[width=\linewidth]{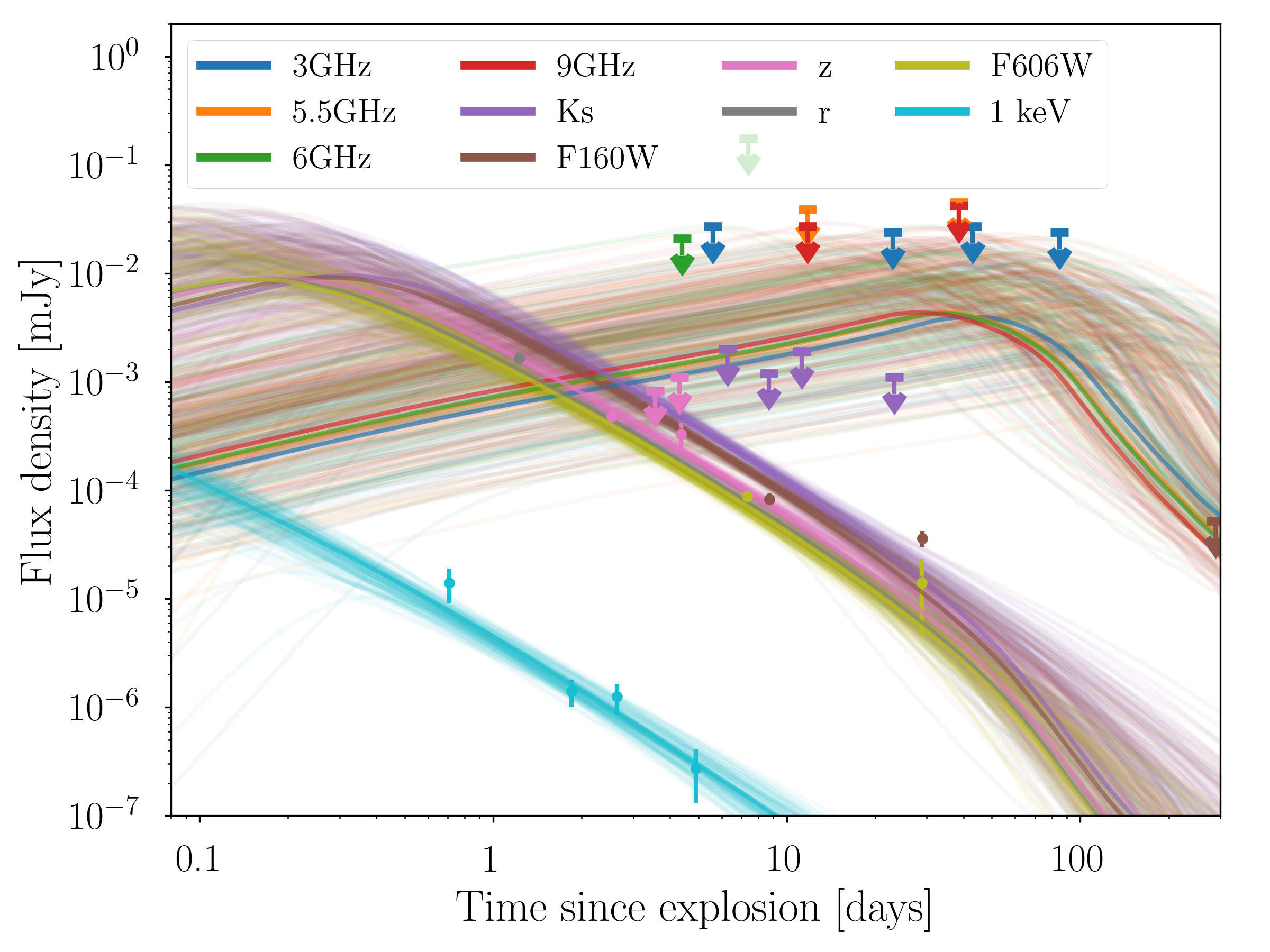}
    \caption{\textsc{Redback tophat} afterglow model fit to the X-ray, optical, NIR
    and radio observations at a fixed redshift of $z=3.2$. The last epoch of HST F606W is assumed to be host-galaxy light only and this flux density has been subtracted from the shown F606W measurements. The fit uses the \texttt{nessai} nested sampler with a Gaussian likelihood.}
    \label{fig:redback_fit}
\end{figure}

We used \textsc{Redback} \citep{redback} to model the radio, optical, near-infrared and X-ray light curves with the \texttt{tophat\_redback} afterglow model at the $z=3.2$. We model the data presented in this work and all the data presented in \cite{Jonker2026} together with additional radio, X-ray and $K_s$-band data and upper limits from \cite{Becerra2026}.
This model is based on \cite{Lamb2018}. We employed the \texttt{nessai} nested samples \citep{nessai} through the \textsc{bilby} framework \citep{bilby}. We assumed a standard Gaussian likelihood, modified such that it can handle upper limits. The best fit is shown in Fig.~\ref{fig:redback_fit} and the corner plot with the posterior distributions is shown in Fig.~\ref{fig:redbackcorner}. We show the fitted parameters, including their priors and posteriors in Table~\ref{tab:redback_parameters}. We also obtained a fit including the dust attenuation parameter, but we find that the changes in the other parameters are well within the uncertainties, and we therefore present the fit with the least parameters, i.e., without dust attenuation.
From the fit, we find a jet energy of $E_0$~=~6.2$^{+35}_{-5}~\times~10^{51}$~erg observed on axis, with a jet core angle of $14^{+2}_{-5}$~degrees and observer angle of $11^{+2}_{-4}$~degrees, hence we observe the jet on-axis (our jet model being a tophat model). 
We conclude that the multi-wavelength data of EP250207b can be described well by a typical GRB afterglow \citep[e.g.,][]{Zhang2006}. 
The interstellar medium number density log($n_{ism}$) is relatively low and therefore more in the range of short GRBs \citep[e.g.~][]{Fong_2015}, although the posterior is wide enough (Fig.~\ref{fig:redbackcorner}) that it also overlaps with the ISM number density typically seen for long GRBs \citep[$n\sim1$~cm$^{-3}$, e.g.~][]{Chrimes_2022}.

\begin{table}
    \centering
        \caption{Free parameters and their priors and posteriors of the \textsc{Redback} modelling using the \texttt{tophat\_redback} afterglow model. The priors are uniform (in log space), unless stated otherwise.}
    \begin{tabular}{c c c }
    \hline \hline
        Parameter & Prior & Posterior\\ \hline
        Viewing angle $\theta_{\rm{observer}}$& Sine (0,$\pi$/2) & 0.19 $\pm$ 0.05 \\
        Jet energy E$_{0}$ & $\log$(44, 54 [erg]) & 51.2$^{+0.8}_{-0.6}$\\
        Jet opening angle, $\rm{thc}$ & (0.01, 0.1) & 0.24$^{+0.04}_{-0.08}$\\ 
        ISM number density $n_{\rm{ism}}$ & $\log$(-5, 2) & -1.0$^{+1.1}_{-1.4}$\\
        Electron power law index $p$ & (1.4, 3.2) & 2.8$^{+0.2}_{-0.3}$\\
        Partition fraction in electrons $\epsilon_e$& $\log$(-5, 0) & -0.3$\pm$0.2\\
        Partition fraction in magnetic field $\epsilon_b$ & $\log$(-5, 0) & -2.4$^{+0.8}_{-1.0}$\\
        Initial Lorentz factor g$_0$& (40,1000) & 560$\pm$300\\ 
        \hline
    \end{tabular}
    \label{tab:redback_parameters}
\end{table}

\subsection{Host galaxy}
We use SE to measure the centroid position of EP250207b on a subtraction image between the first and third epoch of the HST F160W images, and that of the host galaxy by measuring the centroid position in only the third epoch. We find an offset between the two of 0.07$\pm$0.001\arcsec, i.e.~one pixel, which corresponds to 0.5$\pm$0.01~kpc at $z=3.2$. The location of EP250207b is therefore fully consistent with being on top of the newly identified host galaxy. Such an offset is consistent with both collapsar and merger-driven GRBs \citep{Bloom2002,Blanchard2016}. The galaxy has an absolute magnitude of $M_{F160W} = -18.6$.
From the HST F606W image we measure a half-light radius of the host galaxy of 1.7$\pm$0.1 pixels or 0.056$\pm$0.003\arcsec, corresponding to 0.43$\pm$0.03~kpc at $z=3.2$ and from F160W we measure a half light radius of 1.98$\pm$0.5 pixels, which corresponds to 0.13$\pm$0.04\arcsec or 1$\pm$0.3~kpc. These were obtained using the \texttt{FLUX\_RADIUS} parameter in SE, and the quoted uncertainties were estimated by varying the \texttt{DETECT\_THRESH} parameter from 1$\sigma$ to 2.5$\sigma$ in steps of 0.5$\sigma$ and taking the resulting spread in the half-light radii. The half-light radii measured in both filters imply that this is a highly compact galaxy. This compactness is also seen in several collapsar GRB host galaxies \citep[e.g.~table A1 in][]{Lyman2017}. For short GRBs such compact hosts are rare, but there are a few cases of short GRBs associated to such compact dwarf galaxies \citep[][]{Nugent2024}. Based on the offset measured in the F160W images and the half light radius in this filter, we find a host-normalised offset of 0.5$\pm$0.2.

\begin{figure*}
	\includegraphics[width=0.9\textwidth]{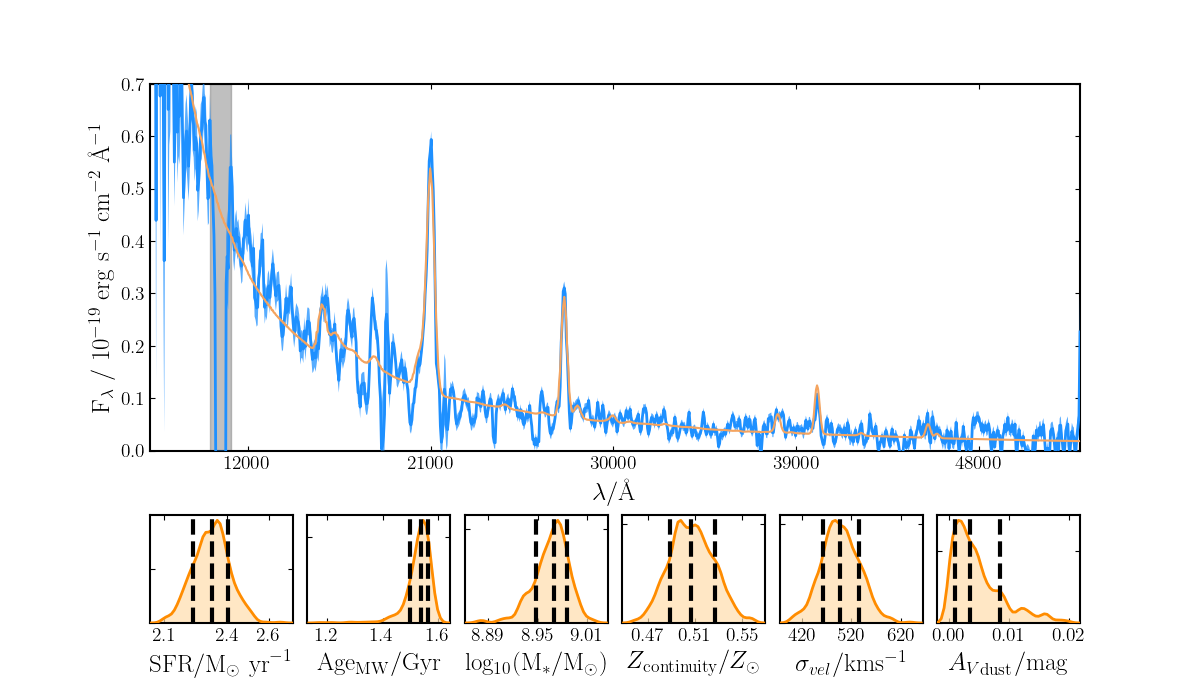}
    \caption{\textit{Top panel:} BAGPIPES fit (orange) to the host galaxy spectrum (blue) in $F_{\lambda}$ in $10^{-19}$~erg~s$^{-1}$~cm$^{-2}$~\AA$^{-1}$. The 1$\sigma$ uncertainty on the spectrum is shown by the blue shaded region. The grey shaded region has been masked during the fit. \textit{Bottom panels:} Posterior distributions for six fitted parameters: SFR, age, stellar mass, metallicity, velocity dispersion and dust extinction. The vertical dashed lines indicate the 16th, 50th and 84th percentiles.}
    \label{fig:BP}
\end{figure*}

We fit our 1D extraction of the JWST/NIRSpec IFU spectrum using the Bayesian Analysis of Galaxies for Physical Inference and Parameter Estimation \citep[BAGPIPES;][]{bagpipes} code to infer properties of the galaxy. As star formation history (SFH) model we considered the \cite{leja2019} continuity non-parametric model with Student's t-distribution priors (-10, 10) for the star formation in each age bin. The edges of the eight age bins are logarithmically spaced according to the age of the universe at $z=3.2$, resulting in bin edges of (0, 10, 24.1, 58.2, 140.5,  339, 818, 1700.9, 1950)~Myr. Furthermore, we adopt free parameters for the formed stellar mass, the metallicity, the \cite{Calzetti2000} dust parameter $A_V$, the velocity dispersion and we set a prior on the redshift of the galaxy (3.15, 3.25). The best fit and the posteriors of six of the free parameters are shown in Fig.~\ref{fig:BP}, where the mass-weighted age (Age$_{MW}$) and the star formation rate (SFR) are determined by the SFH model. We do not show the redshift posterior as it is narrowly distributed around $z=3.2$. 
From the fit we obtain an SFR of 2.3$\pm$0.1~M$_{\odot}$~yr$^{-1}$ and stellar mass of $\sim10^{9}$~M$_{\odot}$, in line with the properties for the host galaxies of both long and short GRBs \citep[e.g.,][]{Li_2016,Nugent_2022}. We compare the present-day specific SFR $\mathrm{log(sSFR)}~=~-8.60~\pm~0.03~\mathrm{yr}^{-1}$ to the short GRB host galaxy sample of \cite{Nugent_2022} and find that it is higher than $\sim86\%$ of the sSFRs in their sample, and to the long GRB host galaxy samples of \cite{Svensson2010, Wang2014, Niino2017} and find that this sSFR is higher than $\sim71\%$ of the long GRB host galaxies. Given the redshift of this galaxy, a higher sSFR can be expected than for the better-studies lower redshift sample. \cite{Hunt2014} find several long GRB hosts with similar sSFR for redshifts $z\gtrsim1$. 

The metallicity of Z$_{*}= 0.51~\pm~0.02$~Z$_{\odot}$ is also well within the range observed for both short and long GRBs \citep[e.g.,][]{Nugent_2022,Palmerio2019}, but we note that metallicities obtain through SED fitting are not suitable to directly compare to those obtained through measurements of line ratios as they measure the stellar metalicity rather than that of the gas. Hence, we additionally compute the gas-phase oxygen abundance metallicity ($12~+~\mathrm{log}(O/H)$) with the emission line fluxes from the best-fit BAGPIPES model. We follow the method described in \cite{Kewley2019} that uses the [NII]$\lambda$6584/[SII]$\lambda$6717.31 ratio and the ionization parameter $U_{gas}$. We cannot use the method that depends on $R_{23}$ as the resulting metallicity from this method is lower than the range where this method is valid \citep[12~+~log(O/H)~<~8.53;][]{Kewley2019}. The resulting gas-phase metallicity is 12~+~log(O/H)~=~8.28~$\pm$~0.03, or log$(Z_{gas}/Z_{\odot})~=~-0.41~\pm$~0.01, where the uncertainties are based on the 16th and 84th quantiles of the posteriors on the line fluxes. This metallicity is in line with the low metallicities observed for long GRB host galaxies \citep[e.g.,][]{Wang2014, Palmerio2019, Schady2024} and similar, although somewhat lower, than for short GRBs \citep[e.g.~][]{Berger2009}.

Considering all properties of the host galaxy, it seems more consistent with the host galaxies of collapsar GRBs than with those of short GRBs,  although the latter cannot be ruled out. We note that is this galaxy is likely more compact and more actively star-forming than the host galaxies typically studied at lower redshift \citep[][]{Taggart2021}.

\section{Conclusion}
\label{conclusion}
We present deep HST and JWST observations revealing an underlying galaxy at the position of EP250207b. \cite{Jonker2026}  and \cite{Becerra2026} had suggested a merger origin for this FXT, due to its proximity on the sky to a galaxy at $z=0.082$, and the non-detection of a supernova at a brightness expected at that redshift. The spectroscopic redshift of an underlying galaxy is $z=3.2$, which we propose is the actual host and thus redshift of the transient. However, for a transient source at this newly found redshift we can no longer rule out a collapsar origin for EP250207b.
 
\textsc{Redback} modelling of the afterglow with a tophat jet model provides best-fit parameters consistent with those of a typical GRB afterglow observed on-axis. The specific star formation rate, gas-phase metallicity and compactness of the underlying galaxy seem to favour a collapsar GRB origin, although a merger-driven GRB cannot be ruled out. We conclude that at this redshift we cannot differentiate between a binary neutron star merger origin and a collapsar origin for EP250207b.

\section*{Acknowledgements}
A.P.C.H., P.G.J., J.Q.V., J.N.D.D., and J.S.S.~are supported  by the European Union (ERC, Starstruck, 101095973, PI Jonker). Views and opinions expressed are however those of the author(s) only and do not necessarily reflect those of the European Union or the European Research Council Executive Agency. Neither the European Union nor the granting authority can be held responsible for them. 
F.E.B/~acknowledges support from ANID-Chile BASAL CATA FB210003 and FONDECYT Regular 1241005, and JWST-GO-06970.002.
T.-W.C.~acknowledges financial support from the Yushan Fellow Program of the Ministry of Education, Taiwan (MOE-111-YSFMS-0008-001-P1), and from the National Science and Technology Council, Taiwan (NSTC 114-2112-M-008-021-MY).
G.P.L.~is supported by a Royal Society Dorothy Hodgkin Fellowship (grant Nos. DHF-R1-221175 and DHF-ERE-221005).
D.B.M.~is funded by the European Union (ERC, HEAVYMETAL, 101071865). The Cosmic Dawn Center (DAWN) is funded by the Danish National Research
Foundation under grant DNRF140.
F.O.~acknowledges support from the INAF-Large Grant 2024:"Envisioning Tomorrow: prospects and challenges for multimessenger astronomy in the era of Rubin and Einstein Telescope"; the INAF-GO Large Grant: "Exploitation of optical and near-infrared followup data of Gamma-ray Bursts" and the INAF-MINIGRANT (2023): "SeaTiDE - Searching for Tidal Disruption Events with ZTF: the Tidal Disruption Event population in the era of wide field surveys". 
M.E.R.~received the support of the Junior Leader Fellowship from ”la Caixa” Foundation (ID 100010434). The fellowship code is LCF/BQ/PI25/12100030.
D.M.S.~acknowledges support through the Ramón y Cajal grant RYC2023-044941 funded by MCIU/AEI/10.13039/501100011033 and FSE+.

This work is based in part on observations made with the NASA/ESA/CSA James Webb Space Telescope and the NASA/ESA/CSA Hubble Space Telescope. The data were obtained from the Mikulski Archive for Space Telescopes at the Space Telescope Science Institute, which is operated by the Association of Universities for Research in Astronomy, Inc., under NASA contract NAS 5-03127 for JWST and NAS 5–26555 for HST. These JWST observations are associated with program \#06970 and the HST observations are associated with program \#17806. 

The MeerKAT telescope is operated by the South African Radio
Astronomy Observatory, which is a facility of the National Research
Foundation, an agency of the Department of Science and Innovation.
This work has made use of the “MPIfR S-band receiver system”
designed, constructed and maintained by funding of the MPI für
Radioastronomy and the Max Planck Society.

\section*{Data availability}
The data underlying this article are available in the MAST archive
and can de accessed via \url{https://doi.org/10.17909/bk0y-h411}.

\bibliographystyle{mnras}
\bibliography{bibliography}

\appendix
\section{Radio observations}
\begin{figure}
    \centering
    \includegraphics[width=\linewidth]{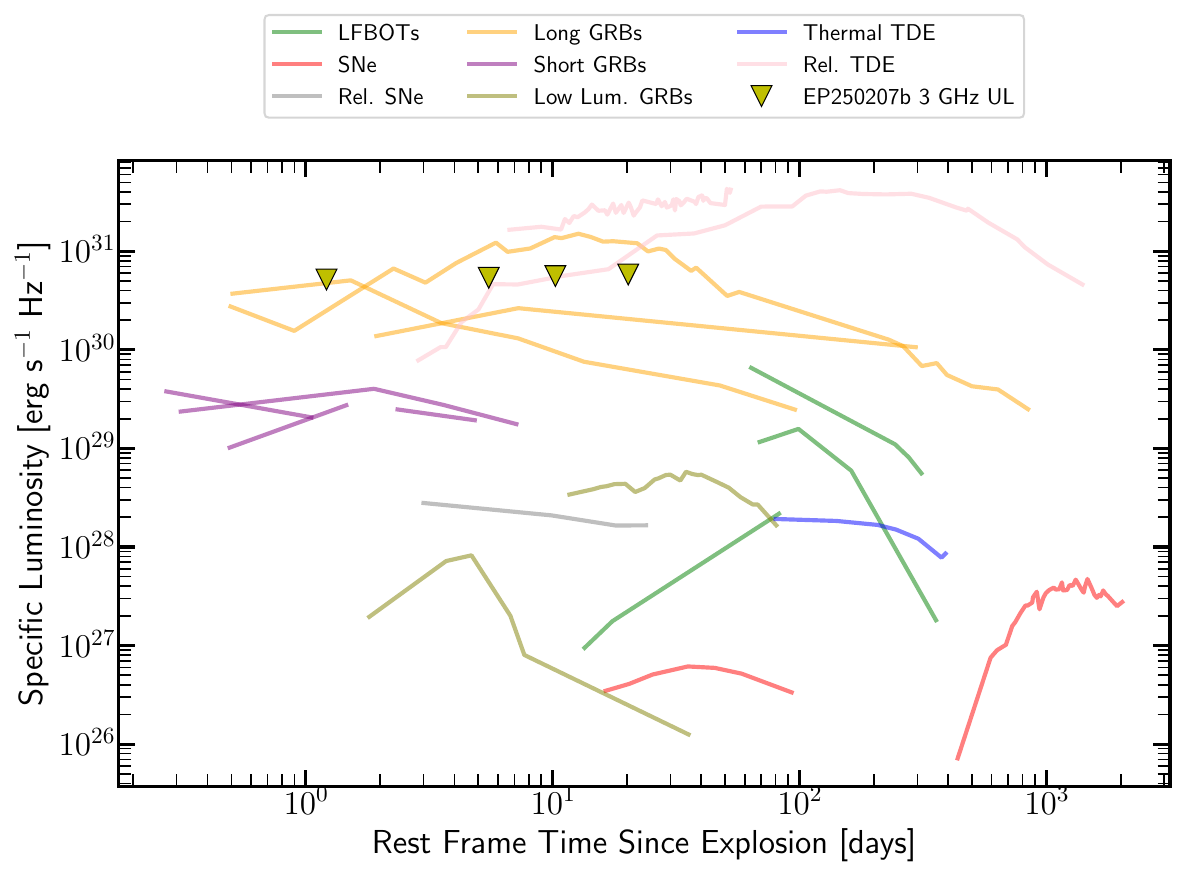}
    \caption{Radio luminosity versus rest-frame time since the explosion for different transient classes. We show the four upper limits in the 3.06~GHz band of EP250207b with triangles for $z=3.2$.}
    \label{fig:radio}
\end{figure}

We show the specific radio luminosity versus rest frame time since explosion of different transient types in Fig.~\ref{fig:radio}

\section{Afterglow modelling posterior distribution}
\begin{figure*}
    \centering
    \includegraphics[width=0.9\textwidth]{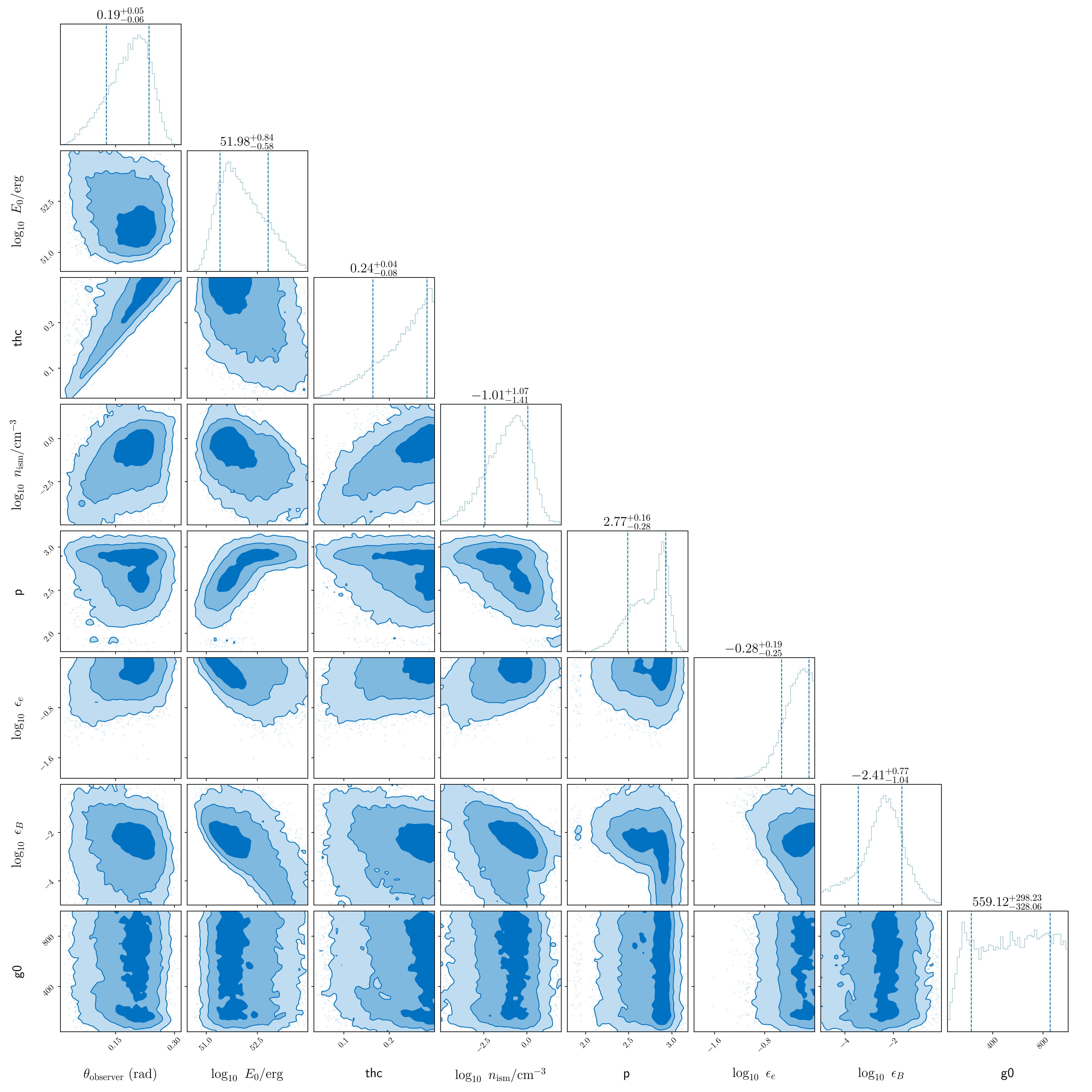}
    \caption{Corner plot of the posterior distributions for the \textsc{Redback} tophat afterglow model fitted to the radio, optical and X-ray observations of the afterglow of EP250207b.}
    \label{fig:redbackcorner}
\end{figure*}

We show the corner plot of the fitted parameters from our \textsc{Redback} modelling in Fig.~\ref{fig:redbackcorner}.

\bsp	
\label{lastpage}
\end{document}